\documentclass[prb,noshowpacs,showkeys]{revtex4}
\usepackage{amssymb,amsmath}
\pdfoutput=1
\usepackage{graphicx,float}
\usepackage{bm}
\usepackage{array}
\usepackage{algorithm}
\usepackage{algorithmic}

\begin{document}


\title{\bf Fundamental interactions in self-organized critical dynamics on higher-order networks
}
\author{ Bosiljka Tadi\'c$^{1,2,\star}$ and Roderick Melnik$^{3,4,\diamond}$ }
\vspace*{3mm}
\affiliation{$^1$Department of Theoretical Physics  Jo\v zef Stefan
  Institute,  Ljubljana, Slovenia }
\affiliation{$^2$Complexity Science Hub, Josephstaedterstrasse 20,
Vienna, Austria}
\affiliation{$^3$MS2Discovery Interdisciplinary Research Institute; M3AI Laboratory and Department of Mathematics, Wilfrid Laurier University;  Waterloo  ON Canada}
\affiliation{$^4$BCAM - Basque Center for Applied Mathematics; Alameda de Mazarredo 14; E-48009 Bilbao Spain} 
\vspace*{3mm}


\begin{abstract}
\noindent
In functionally complex systems, higher-order connectivity is often revealed in the underlying geometry of networked units.
   Furthermore, such systems often show signatures of self-organized criticality, a specific type of non-equilibrium collective behaviour associated with an attractor of internal dynamics with long-range correlations and scale invariance, which ensures the robust functioning of complex systems, such as the brain.
Here, we highlight the intertwining of features of higher-order geometry and self-organized critical dynamics as a plausible mechanism for the emergence of new properties on a larger scale, representing the central paradigm of the physical notion of complexity. Considering the time scale of the structural evolution with the known separation of the time scale in self-organized criticality, i.e., internal dynamics and external driving, we distinguish three classes of geometries that can shape the self-organized dynamics on them differently. 
We provide an overview of current trends in the study of collective dynamics phenomena, such as the synchronization of phase oscillators and discrete spin dynamics with higher-order couplings embedded in the faces of simplicial complexes.
For a representative example of self-organized critical behaviour induced by higher-order structures, we present a more detailed analysis of the dynamics of field-driven spin reversal on the hysteresis loops in simplicial complexes composed of triangles.
These numerical results suggest that two fundamental interactions representing the edge-embedded and triangle-embedded  couplings
must be taken into account in theoretical models to describe the influence of higher-order geometry on critical dynamics.\\[3pt]
\end{abstract}
\maketitle
\section{Geometry, interactions and emergent behaviour in complex systems\label{sec:intro}}
The role of self-organised criticality (SOC) has been increasingly recognised in various complex systems, from brain functioning to social and geophysical phenomena
\cite{SOC_we_dynamics2021} as a mechanism enabling robust functioning and emergent properties at a larger scale that reside on
collective dynamic behaviours; for a recent review of general features
of complex systems, see
\cite{CS_bookST2018,ComplexSystems_review2020,ComplexSystems_Estrada2023,Noextensive_review2023Sym}. These are nonlinear dynamical systems
repeatedly driven by environmental forces that self-organise
towards an attractor with a stationary state characterised by
avalanches, long-range correlations and scale invariance. These
stationary out-of-equilibrium critical states enable the system's
response to driving forces at all scales, thus providing robust
functioning and stability. Critical dynamics out of equilibrium is
currently an active research field. Two challenging  issues are the impact of a dynamically changing environment and underlying complex geometry \cite{SOC_we_dynamics2021}.

Besides numerical simulations, the renormalisation-group (RG) theory with scaling concepts and methodologies based on quantum and statistical field theory for
equilibrium phase transitions are extended to non-equilibrium dynamics
and SOC; see recent reviews \cite{RG_nonequilibrium_UTauber2017} and
\cite{RG_envirDyn2023Antonov} and references there. The described models with SOC behaviours \cite{RG_envirDyn2023Antonov} demonstrate the field-theory methodologies as powerful tools to
characterise collective fluctuations in stationary states of complex
systems that are driven out of equilibrium. Beyond external and internal noise terms, this methodology properly considers the dynamical environment, which can critically impact the intrinsic SOC dynamics
\cite{antonovSOC2023}. 
Another promising way to understand the role of the complex systems' coupling to the environment can be built by the use of quantum formalism; a representative example is the study of adaptive complex systems \cite{CS_emergentQuantumTh2023}, where the formal theory leads to a requirement that the emergent ``quantum'' potential needs to be effectively balanced by the environmental coupling. It should be noted that the RG theory is based on continuous field models. For example, studies
\cite{antonovSOC2023,hwa1992avalanches,tadic1998disorder} of SOC
are based on continuous versions of the original discrete sandpile
automata (SPA) model, incorporating its essential intrinsic anisotropy
of diffusion on otherwise homogeneous space. However, considering
more complex underlying geometries \cite{Dyn_criticalonNets2008SD} within the RG theory \cite{RG_Networks2018Boguna,Nets_LowDim2022}, in particular, those enabling geometry-embedded higher-order interactions
\cite{HOC_review2023boccaletti,HOC_nonlocal2023Kuehn,we_PhysA2015,HOC_evolution2021}, remains a challenging problem. Here, we aim to highlight the interplay of higher-order geometry and emergent SOC behaviour in complex systems.

Three types of underlying geometry that can shape the collective
dynamics and SOC can be recognised as networks with fixed, co-evolving, and temporal structures. The dynamical units
associated with the network's nodes interact via the network-provided connections of different orders. As explained below, this separation is conditional in real-world systems,  relating the time scale
of the network's evolution vs the dynamics of interacting units on it.

\textit{Fixed geometry network} substrates extend the idea of 
physics models, e.g., spins at sites of a regular 2D- or 3D-lattice
to a more complex structure  described by the network or
Simplical complex measures; see more details below. The dynamical units
associated with the network's nodes and interacting through the
local or nonlocal structure may have identical dynamics, i.e., as phase
oscillators \cite{SYNCH_Arenas_NatureComm2020,Synch_clustersEVlocaliz2023}
or spins on simplicial complexes \cite{SC_we_Entropy2020}; another option often used in the agent-based modelling is that each unit has individual dynamics and parameters characteristic of each agent associated with a given node \cite{OSN_weEntropy2013ABM}. The temporal evolution of each unit is subject to interactions (of different orders), driving fields, and constraints by the surrounding geometry. It can be described by a set of update rules, as in the
case of cellular automata or spin kinetics or by solving differential
equations with identical, for example, in the case of phase
oscillators, or individual forms and parameters, as in the
case of agent-based models.

\textit{Co-evolving networks} represent structural patterns that appear
 through the dynamics of interacting units and develop over time. They 
 can be visualised as mathematical graphs, similar to the OSN of users mentioned above, or by bipartite graphs, for example, in blog
 data, where indirect interactions among users (as one
 partition) are effectuated over another partition as posted subjects
 of communications;  In this
 case, graph mapping enables advanced graph theory methods
 to quantitatively study the structure of these dynamical patterns
 and how they change over time. However, the co-evolutionary
 mechanisms are more subtle, involving collective dynamics with SOC signatures;
see, for
 example,  \cite{OSN_weEntropy2013ABM} and references there.
The geometry emerges from the dynamics of interacting units, meanwhile, the currently existing structure facilitates the diffusion (of information, knowledge, emotion) to spread further. Unsurprisingly, in cooperative users' activities, such as collective knowledge building \cite{dankulov2015dynamics,SOC_wePRE2017knowe}, some stable network structures emerge in the other partition. For example, they represent the
 network of emergent knowledge (of subjects) in the case of knowledge-creation dynamics in Mathematics data or a stable ``social graph''  emerges among users of Ubuntu Chats that persists
 over the years, satisfying the social 'weak ties' hypothesis; see \cite{SOC_weHBNets2019} and references there. 
Evolutionary processes, where the respective time scale is system
characteristic, comprise one of the fundamental features
of complexity \cite{ComplexSystems_review2020}. In this context, the network mappings of the Brain
represent different types of functional patterns rather than a fixed
structure graph;  for a recent survey, see
\cite{Brain_CNdyn_review2024}. A more detailed description of the
human connectome is given in \ref{sec:SOC} .

\textit{Time-varying geometries} represent another category of structures compared to the above-discussed cases; they exhibit partial or global reconstruction, which is virtually independent of the processes of dynamical units at their nodes but occurs at the time scale of these processes.
The structural changes cause \textit{nonlocal effects} that can be examined
directly in real and phase space or indirectly by their
impact on the dynamics. This issue deserves more attention,
particularly in higher-order networks.  In analogy to methods of programable
self-assembly of materials \cite{ProgMatt_Exp2016}, local
reconstruction due to, for example, built-in defects in the simplicial
complexes architecture \cite{SC_we_PRE2020} can cause a  collapse of hierarchical structure at a larger scale.

This perspective paper focuses on the interplay of higher-order geometry features and SOC dynamics. In this context, the time scale of the structural evolution is particularly relevant, given the critical requirement of the time scale separation between the internal dynamics and driving for the occurrence of SOC. In the following, we describe the concepts of SOC and higher-order geometry with simplicial complexes and survey current research trends. As a representative example,  a more detailed study is presented considering spin-reversal dynamics on a structure of geometrically assembled triangles, where the competing pairwise and triangle-embedded interactions lead to emergent SOC behaviours. The paper ends with a brief discussion with a summary of open questions and research directions.

\subsection{Collective dynamics \& Self-organised criticality in
  Complex Systems\label{sec:SOC}}
 As stated above, the term SOC refers to out-of-equilibrium critical behaviour occurring in a steady state near an attractor of intrinsic nonlinear dynamics \textit{without} apparent phase transition. For a more recent review, see
\cite{SOC_Jensen,aschwanden2013self,markovic2014power} and references there. Such attractors are reached by the system's response to repeated driving by external forces when the driving rate is
slow compared to the time scales of the intrinsic
dynamics. Self-organised dynamics possess characteristic
self-similarity manifested in avalanching behaviours, temporal
correlations and scaling, adequately described by theoretical concepts
\cite{antonovSOC2023} and numerical methods \cite{mcateer201625}. 

Ever since Bak and coworkers introduced it and defined the paradigmatic SPA model \cite{SPA_btw,dhar1990self}, the idea of self-organised dynamics was utilised to understand mechanisms underlying complexity
\cite{bak1995complexity,wolf2018physical,SOC_we_dynamics2021}. 
Signatures of SOC are increasingly found in many complex systems
across the scales in physics \cite{aschwanden2013self}---from nano assemblies to rainfalls \cite{SOC_rainfalls2015}, geophysics and solar activity \cite{smyth2019self}, and biology
\cite{SOC_bialek2011,SOCliving_Munoz2018review}. Furthermore, collective dynamics based on social cooperation represents a specific type of SOC behaviour evidenced by empirical data analysis, for example, in human activity devoted to collective knowledge creation \cite{SOC_wePRE2017knowe} but also in schooling fish \cite{SOC_fish2022romu}. Further examples include bio-social epidemic processes 
\cite{SOC_Epi2000,SOC_Epi2014qexpdengue}, socio-technological 
\cite{SOC_trafficmanagement2023,SOC_Optimization2018}, and
socio-economical systems \cite{SOC_econo2021review}. In addition to
characteristic avalanches superimposed to cyclical trends \cite{we_cycles_emo23}, monitoring changes in the co-evolving network geometry serves as a ``blueprint'' of complexity in social dynamics
\cite{SOC_weHBNets2019}.

\begin{figure}[htb]
\begin{tabular}{cc}
\resizebox{32pc}{!}{\includegraphics{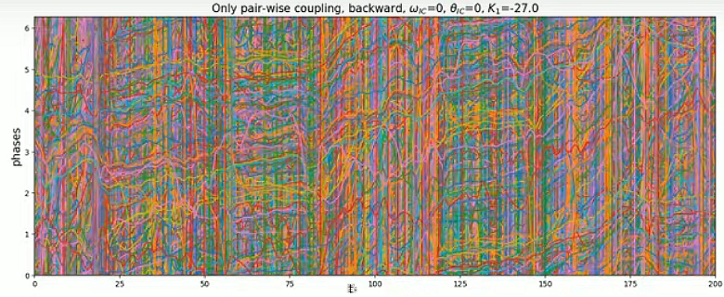}}\\
\end{tabular}
\caption{A pattern of phase evolution at negative pairwise coupling with partially synchronised groups of nodes (brain regions) in human connectome core networks, consisting of simplexes of all orders attached to the eight Brain hubs. Data from \cite{we_Braindyn2022ChSFR}.}
\label{fig:brainsynch}
\end{figure}

Understanding Brain functions and mechanisms of Brain disorders in neurology \cite{Brain_strfunct2023review,Brain_disorders2023imaging,Brain_SOCpsychedelic2023} represent the most challenging issues in the science of complexity and their applications, in particular, in the era of artificial intelligence  \cite{Brain_AI2022dynmodels}.
In this context, network mapping \cite{Brain_horizons2022review} and
the complexity of the Brain dynamics as a SOC system of firing
neurons are central questions both in theoretical and experimental research
\cite{Brain_SOC2014Thilo,gros2021devil,Brain_SOCexp2021review}. 
The whole Brain computational connectomics
\cite{Brain_Connectomics2017review} and imaging techniques
monitoring different  Brain functions, for example, attention \cite{Brain_intseg2021attention} and cognition \cite{Brain_intsegEEG2018cognition} or pain processing
\cite{Brain_intseg2022pain}, reveal multiple scales  interplay between
integration and segregation processes that involve distinct Brain regions and variable communications among them. Therefore, information about the architecture of these distributed neuronal processes can be gained by
modelling Brain dynamics and identifying temporal variations of local order in the
corresponding Brain networks; see a recent survey in 
\cite{Brain_Nets2023review}. 
 Among phenomenological models, synchronisation processes of Kuramoto phase oscillators are often studied on Brain networks.
In this picture,  ``the human brain is a complex system comprising 
 subregions that dynamically exchange information between its various parts through synchronisation'' persisting in a natural metastable at the edge of synchrony \cite{Brain_synch2021psychiatry}. These studies gain some importance in quantifying brain dynamics  in various psychiatric illnesses and Brain disorders implicating altered metastability 
      \cite{Brain_synch2021psychiatry,Brain_Synch2023PLOScbiol}.  
Mathematically, it was shown\cite{Synch_noBrainsyncro2019} that the
structure of the Brain network does not permit a stable
full synchronisation \cite{Synch_noBrainsyncro2019}. Potential mechanisms to maintain partial synchronisation with co-evolving groups of weakly synchronised nodes are demonstrated in the human connectome, as illustrated in Fig.\
\ref{fig:brainsynch}, considering the core network around Brain bubs \cite{we_Braindyn2022ChSFR}.

\subsection{Higher-Order Connectivity: Dynamics on Simplicial
  Complexes \label{sec:SCdyn}}

In networks mapping functional connections in many complex systems,
hidden geometry of higher-order relations occurs \cite{Nets_3xBoguna2021}.
They can be revealed by advanced mathematical techniques beyond standard graph theory \cite{sergey-lectures}, such as the algebraic
topology of graphs \cite{kozlov-book}. They are described by aggregates of simplexes (simplicial complexes) \cite{jj-book} and multigraphs \cite{}. For example, in social graphs \cite{we_PhysA2015}, Human connectome
\cite{we-Brain_SciRep2019,HOC_neuronsnets2019,HC_cliquecavities_JCompNeurosci2018,we-Brain_SciRep2020},
etc., such structures naturally evolved with the self-organised
dynamics. Meanwhile, in materials design \cite{AT_Materials_Jap2016}
such complex structures often emerge from self-assembly processes, particularly those based on preformatted building blocks, e.g., groups of nanoparticles \cite{AT_cooperativeSA2021}.
Such higher-order geometries provide a basis for multiple interactions that play their role in the dynamics and determine the system's collective behaviours. Therefore, studies of the impact of higher-order geometries on the dynamics are vital for understanding the mechanisms underlying dynamic critical behaviours and can be used to estimate the predictability limits \cite{shapoval2021predictability} in the system's evolution. The interplay of the dynamics and higher-order structures can be also utilised to design new methodologies of network control
\cite{HOC_control2023ChSFR}.

Recently, attention has been devoted to the dynamics of units
associated with the nodes in simplicial complexes with a fixed structure
\cite{HOC_review2023boccaletti}. Below, we
describe some key features of such geometries and embedded
higher-order interactions in the case of spin-reversal dynamics and
synchronisation among phase oscillators. For this type of study, an underlying simplicial complex is grown, i.e., using a model; depending on the research aims, several generative models of simplicial complexes with different emergent structures are known in the literature \cite{SC_we_SciRep2018,SC_flavorquantum2016GB,SC_we_PRE2020,SC_Sfboccaletti2021}.
 Alternatively, a real-world network, e.g., human connectome \cite{we_Braindyn2022ChSFR}, can be considered, and methods of Q-analysis\cite{Q-analysis-book1982} applied to determine its detailed architecture \cite{we-Brain_SciRep2019}.

\begin{figure}[htb]
\begin{tabular}{cc}
\resizebox{32pc}{!}{\includegraphics{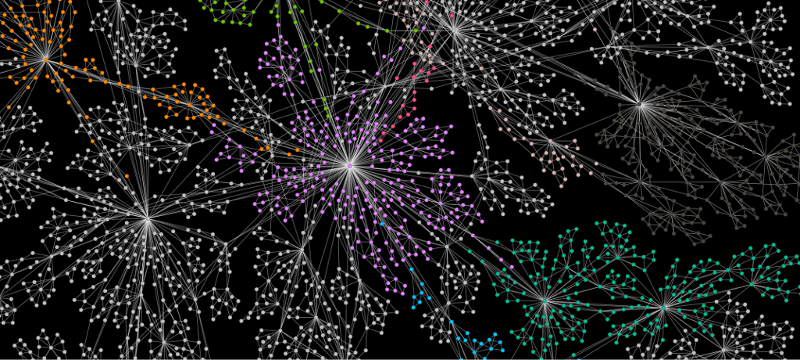}}\\
\end{tabular}
\caption{The simplicial complex of triangles self-assembled under chemical affinity and geometric compatibility rules in (\ref{eq:pattach}) with $\nu=5$;  a close-up view with hubs is shown, and colours indicate mesoscopic communities.}
\label{fig:netstructure}
\end{figure}

Here, we briefly describe the model  introduced in \cite{SC_we_SciRep2018}, where the rules for the attachment of simplexes are motivated by the above-mentioned
cooperative self-assembly of nanostructured materials
\cite{AT_cooperativeSA2021}.  
Specifically, starting with a single simplex of the size $s=q_{max}+1$, at each growth
step $t_g$, a new simplex
is added such that it shares one of its faces of the order
$q=0,1\cdots q_{max}-1$, i.e., a node, an edge, triangle, etc.,  with an existing simplex, randomly selected in the structure; the attaching
probability is given by 
\begin{equation}
P(q_{max},q;t_g)= \frac{c_q(t_g)e^{-\nu (q_{max}-q)}}{\sum _{q=0}^{q_{max}}c_q(t_g)e^{-\nu (q_{max}-q)}} \ .
\label{eq:pattach}
\end{equation}
where $c_q(t_g)$ stands for the number of
geometrically compatible locations that are found in the existing
structure at the growth step $t_g$. The parameter $\nu$ represents the
chemical affinity of the existing structure towards adding of
$q_{max}-q$ new nodes. 
Thus, simplicial complexes with a sparse architecture are grown when
$\nu<0$, representing a ``tree of cliques'' that predominantly share a
single node, whereas the structure becomes increasingly more compact with increasing $\nu>0$, where sharing larger faces is more probable. 
When $\nu=0$, the process is controlled by strictly geometrical
rules. We note that the spectral dimension of the Laplacian operator
associated with the adjacency matrix of the underlying graph, i.e., 
(1-skeleton)  of a simplicial complex is another measure of its
architecture that determines the nature of collective dynamics on
it \cite{Spectra_Nets2003SD,Spectra_dsSynch2019GB}.
The graphs of complexes grown at different chemical affinity
is characterised by the spectral dimension that varies from the 
random-tree  dimension at $\nu<0$ through $d_s\sim 2$ at $\nu=0$, and
continuously increasing with $\nu >0$ and the dimension of simplexes;
see detailed analysis in   \cite{Spectra_wePRE2019SC}. 
The size of the newly added simplex can be fixed in
advance or drawn from a given distribution of sizes; see original reference \cite{SC_we_SciRep2018} for examples and detailed analysis of the architecture of complexes for different chemical affinity and their Q-analysis.   Note that by construction, the distances between the building simplexes are zero, and then these simplicial complexes are
1-hyperbolic \cite{HB_cliques2017plus1}; similarly, 
the size of the largest clique determines the dimension of the simplicial complex; see detailed discussion in \cite{SC_we_SciRep2018}. 
For demonstration, an example of self-assembled triangles is shown in Fig.\ \ref{fig:netstructure}.  It exhibits several branches of triangles attached via shared nodes and edges. These branches often appear to belong to topological communities, here indicated by different colours; they are interconnected through large hub nodes.

A representative structure of the assembled triangles based on geometrical attachment rules is  used in this work in section\ \ref{sec:HLC}.
In this case, we have $\nu=0$, hence  the probability of sharing  faces of the size $s=1$ (nodes) and $s=2$ (edges) are governed by strictly geometric compatibility rules; its spectral dimension is $d_s\simeq 2.1$; see \cite{Spectra_wePRE2019SC}. 
We note that about  21\% of edges and 33\% of single-node faces are shared among two or more triangles in this type of structure.
Moreover, as shown in \cite{SC_HOCweEPL2020}, assortative nodes' correlations are observed and broad distributions of the generalised degree (number of edges per node and the number of triangles per node, which behave statistically similar in this case). The degree distribution shows a power-law decay with an exponent $\gamma \geq 3$ for a segment of intermediate degrees, excluding the nodes with extensive connectivity (hubs). 
During the assembly process we enumerate each edge $\{<i_I,j_I>\}$,
$I=1,2,\cdots E$ and each triangle $\{<i_T,j_T,k_T>\}$, $T=1,2,\cdots \Delta$ and make the lists of the nodes that make them; such lists comprise the corresponding adjacency tensors for the geometry-embedded
interactions among spins, as discussed in the following section.
In the remaining part of this section, we give an insight into how the
structure of simplicial complexes yields the geometry-embedded
interactions in the more general cases and how they are taken into account in numerical
simulations of two fundamental processes leading to collective phenomena: phase synchronisation and spin dynamics.

\textit{Spin dynamics on SC:} We consider the case where  Ising spins $S_i=\pm 1$ is attached to each node, and interactions among spins enabled by the simplexes and their faces of different sizes $k^\prime=2, 3 \cdots q_{max}+1$, where $q_{max}$ is the order of the simplicial complex in question. Given the pairwise adjacency matrix, as is usually the case with real-world network data, all simplexes and their faces can be identified by Q-analysis, as
stated above. Alternatively, in the self-assembly model described
above, by monitoring all added simplexes, we keep track of the
identity of the nodes that make them. In this way,  we can produce a unique list of simplexes of all sizes $k^\prime$; then, we can write the Hamiltonian  with all possible geometry-embedded interactions in that simplicial complex as follows \cite{SOC_we_dynamics2021}
\begin{equation}
{\cal{H}} = -\sum_{k^{\prime}=2}^k\sum_{(i_1,\cdots
  i_{k^{\prime}})\in{\cal{L}}_{k^{\prime}}}J_{i_1,i_2,\cdots i_{k^{\prime}}}S_{i_1}S_{i_2}\cdots S_{i_{k^{\prime}}}
-h\sum_iS_i  \ .
\label{eq:Ham}
\end{equation}
Here, $h$ is the external magnetic field and $J_{i_1,i_2,\cdots
  i_{k^{\prime}}}$ stands for the interaction tensor of the order
$k^{\prime}$, whose elements differ from zero  when the
indexes $(i_1,i_2,\cdots  i_{k^{\prime}})$ match one of the simplexes in the list ${\cal{L}}_{k^{\prime}}$ of simplexes of the size $k^\prime$, and zero otherwise. For example, $J_{i_1,i_2}$ is nonzero when the indexes correspond to the nonzero elements of the network's adjacency matrix; similarly, the nonzero values of $J_{i_1,i_2,j_3}$ correspond to all triangles in the structure, and so on. As usual in studies of spin systems with pairwise interactions, the sign and strength of interactions and potential disorder, as well as the presence of given higher-order interactions, are determined considering the physics of the problem and the study objectives.

\textit{Synchronisation on SC:} In most studied Kuramoto model that we present here, 1-dimensional phase oscillators $\theta_i$ are associated with the nodes $i=1,2 \cdots N$ of the simplicial complex,
and the geometry-embedded interactions are provided by simplexes of
different orders $q=0,1,\cdots q_{max}$; see review in \cite{HOC_review2023boccaletti}.
 For the synchronisation of topological signals associated with faces
 of simplexes, see recent work in \cite{SYNC_globalGinestraPRL23} and
 references there. Then, the evolution of phases of all nodes is obtained by solving the differential equations interconnected via interaction tensors, as described below. For this purpose, for each
 interaction order $q$, we prepare a unique list ${\cal{L}}_i^{q}$ of
 the simplexes of the order $q$ that contain the considered node $i$. 
Then, the generalised equation of motion can be written as 
\begin{equation}
\frac{d\theta_i}{dt} = \omega_i -\sum_{q=1}^{q_{max}} \frac{K_q}{qk_i^{(q)}}
\sum_{(j_1,j_2\cdots j_q)\in {\cal{L}}_q^{i}}B_{i,j_1,j_2\cdots j_q}\sin\left(q\theta_i-\sum_{m=1}^q\theta_{j_m}\right) \ 
\label{eq:synch}
\end{equation}
where $\omega_i$ is the node's internal frequency and the interaction tensor $B_{i,j_1,j_2\cdots j_q}=1$ when the set of indexes $(i,j_1,j_2\cdots j_q)$ matches one of the entries in the list ${\cal{L}}_i^{q}$ of simplexes of the order $q$, and zero otherwise;
the number of  such simplexes is the node's $i$ generalised degree
$k_q^{(i)}$. Note that the coupling function in (\ref{eq:synch}) satisfies general conditions for the diffusive-like, non-invasive and natural coupling \cite{HOC_review2023boccaletti}.
The corresponding interaction constants $K_q$ are varied 
to explore the system's transition to the synchronised states; see for
example,  \cite{SC_synchro_wePRE2021,SC_synchro_wePRE2023}, for the case of  high-dimensional simplicial complexes with the pairwise and
triangle-embedded interactions. 
It has been recognised that higher-order interactions cause new collective dynamics phenomena; for example, triangle-embedded interactions induce the broadening of the hysteresis loop and an abrupt
desynchronisation transition \cite{SYNCH_Arenas_NatureComm2020}.
Studies \cite{Spectra_dsSynch2019GB,SC_synchro_wePRE2023} have demonstrated the relevance of the architecture of simplicial complexes to synchronisation, quantified by varied spectral dimensions. Moreover, spectral analysis and eigenvector localisation have recently been explored to predict cluster synchronisation in theoretical \cite{Synch_clustersEVlocaliz2023} and  experimental
studies \cite{Synch_groupsPRE2013experiment,Synch_clusters2020NatComm}.

\section{Hysteresis-loop self-organised criticality with triangle-embedded interactions\label{sec:HLC}}
As mentioned above, here we present a more detailed analysis 
of the spin-reversal dynamics driven by the slow ramping of the magnetic field along the
hysteresis loop, demonstrating the emergence of SOC due to complex
geometry. The spins are situated at nodes of a large complex of self-assembled triangles; cf.\  Fig.\ \ref{fig:netstructure}.  Therefore, we have two types of geometry-embedded interactions in the Hamiltonian (\ref{eq:ham}), specifically:
\begin{equation}
{\cal{H}} = (\kappa -1)\sum _{(i,j)\in {\cal{L}}_2}J_{ij}S_iS_j
-\kappa\sum_{(i,j,k)\in{\cal{L}}_3}J_{ijk}S_iS_jS_k- h\sum _iS_i  \ ,
\label{eq:ham}
\end{equation}
where ${\cal{L}}_2$ and ${\cal{L}}_3$ stand for unique lists of the network's edges and triangles, respectively. 
For this study, we set $J_{ij}=-J_2$ fixing the antiferromagnetic pairwise interactions among all implicated pairs of spins, and  $J_{ijk}=J_3$; a parameter $\kappa
\in [0,1]$ is added to balance their respective contributions. 
As shown below, we can differentiate the ferromagnetic and  $J_{ijk}=-J_3$ for antiferromagnetic triangle-embedded interactions; see also \cite{SC_HOCweEPL2020}.
As usual,  the
dimensionless units can apply; thus, we set $J_2=1$ and
$J_3=1$. Moreover, the magnetisation
(in Bohr magnetons $\mu_B$) is determined as $M=(N_+ -N_-)/N \in
[-1,1]$ by the balance of the respective up and
down oriented spins $N_+$ and $N_-$ with respect to their number  $N$. The
external field (in the units
$\mu_B$) , $h\in [-h_{max},+h_{max}]$, where $h_{max}$ is
related to the maximum number of neighbours of the vertices,  is varied in a quasistatic manner, as explained below.

Starting with a uniform state of all spins $\{S_i=-1\}$ and
$h=-h_{max}$, the spin-reversal process is driven by slow field ramping $h\to h+\delta h$ along an
ascending  branch of the hysteresis and then reversing the field to
close the loop. As it is a widely accepted approach in the study
of Barkhausen noise in disordered magnetic systems a zero-temperature dynamics is applied; see, for example,  \cite{tadic2019critical} and references
therein. 
In particular,  the spin $S_i$ at the node $i$ flips to align along the external field 
which can balance the local field due to the
neighbouring spins, i.e., when $h_i^{loc}S_i <0$. Here,
$h_i^{loc}=  (\kappa -1)\sum _{j\in {\cal{L}}_2^i}J_{ij}S_j
-\kappa\sum_{(j,k)\in {\cal{L}}_3^i}J_{ijk}S_jS_k- h$ is the local field
acting on the spin at node $i$ at a current value of the external field $h$ and the summation indicates the  corresponding subsets of the edges ${\cal{L}}_2^i$ and triangles ${\cal{L}}_3^i$ that contain the node $i$.
Thus, the spin-flip at node $i$ causes changes in the
local fields of its neighbours, which can satisfy the condition to
flip, and so on, resulting in an avalanche. 
The avalanche stops when no more spins satisfy the above condition to flip. 
The boundary between the domain of flipped and unflipped spins defines the position of the domain wall at a given value of the external field.
When the avalanche stops, the external field is changed again by $\delta h$ (adiabatic driving). 
We note that spin frustration \cite{Frustr_book2015Julich} with the antiferromagnetic pairwise interaction among spins on a triangle prevents all three spin pairs from simultaneously ordering with the field; thus, some spins may flip back even though the above
condition is fulfilled. To avoid frequent back-and-forth flips of
the same spin at a given field value, flips with a probability $p\lesssim 1$ are adopted. Moreover, without the magnetic disorder, the field ramping parameter $\delta h=1$ is the lowest value that may move the domain wall \cite{SC_HOCweEPL2020,SOC_we_dynamics2021}.

In the simulations, a time step $t$ consists of a parallel update of
all spins in the network. At each time step during the spin activity avalanches, we sample the value of
the external field $h_t$, the total number of flipped spins $n_t$, and the number of spins ordered with the field (unnormalised
magnetisation) $M_t=n_t^{+}-n_t^{-}$. 
Here, our focus is on the distributions of avalanches; for three representative values of the parameter $\kappa$, in particular,  $\kappa =0$, describing purely antiferromagnetic pairwise $-J_2$ interactions, $\kappa =1$,  corresponding to the case where only triangle-embedded interactions $J_3$ are present, and the case $\kappa=0.5$, where these two interaction types are well balanced. 
 (See ref  \cite{SC_HOCweEPL2020} for the analysis of hysteresis loop shape and temporal correlations of the signals $\{n_t\}$.)
\begin{figure}[htb]
\begin{tabular}{cc}
\resizebox{24pc}{!}{\includegraphics{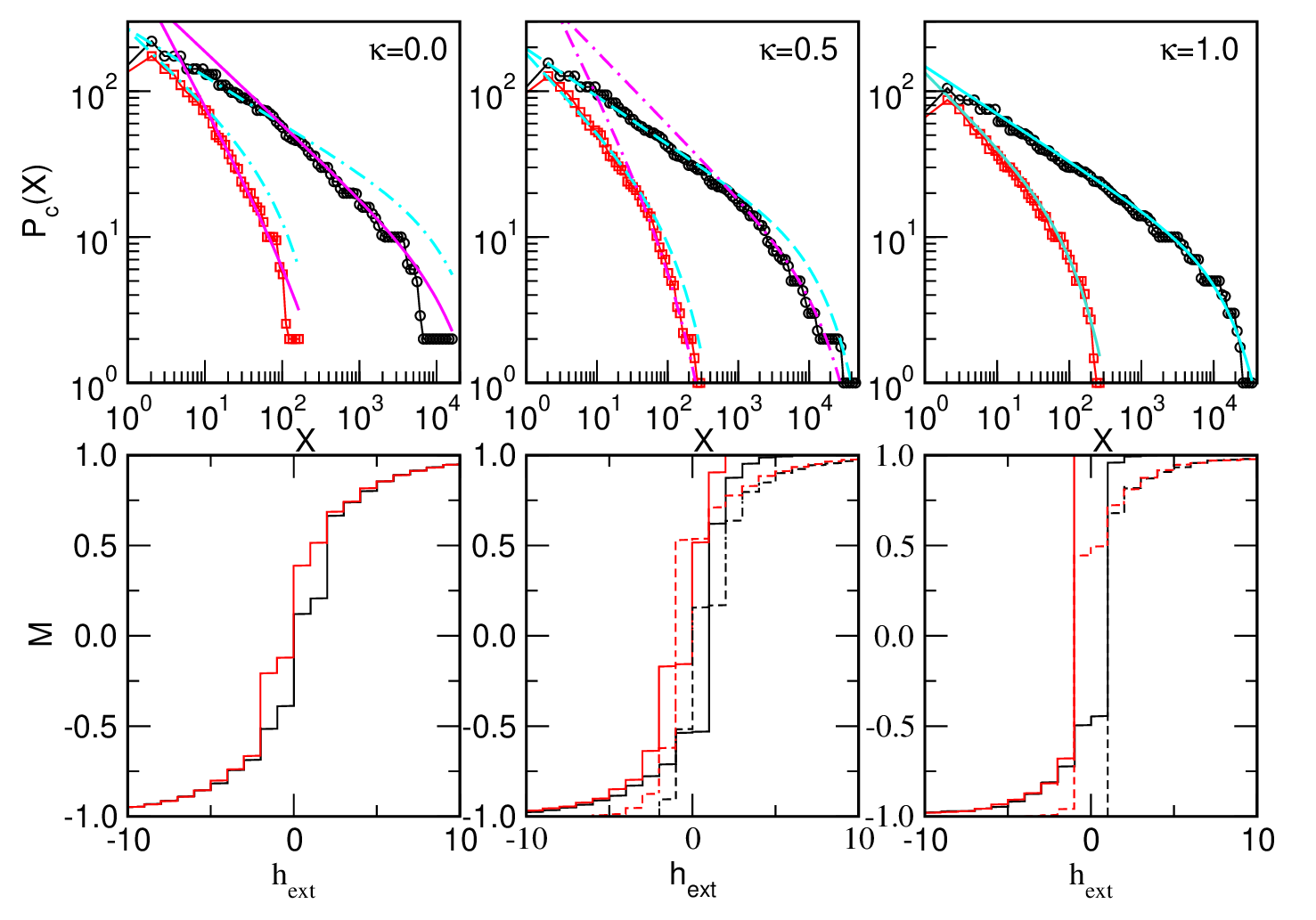}}\\
\end{tabular}
\caption{Top panels: Cumulative distributions of the avalanche size (black
  circles) and duration (red squares) averaged over the entire
  hysteresis loop for $\kappa=$ 0.0, 0.5, and 1.0, as indicated on the
  panels. For the case $\kappa=0.5$, the distributions for $J_3<0$ are
  shown; see text for a detailed description of fits. Lower panels: Hysteresis loops corresponding to the values of $\kappa$ in the panel above; forward/backward branch shown in black/red; full and dotted lines for $\kappa=0.5$ and $\kappa=1$ are for $J_3>0$ and $J_3<0$, respectively. }
\label{fig:distributions3x}
\end{figure}

As this figure shows, the spin-activity avalanches appear in different sizes and durations that can be described by a cumulative distribution function having a power-law segment and an exponential cut-off
\begin{equation}
P_c(X)=AX^{-\tau_X-1}\exp{-X/X_0} \ ,
\label{eq:Pcsizedur}
\end{equation}
 where  $X$ stands for the size $s$ (the number of spins undergoing
dynamics) and the duration $T$ (measured by the number of time steps from
the field rump till the activity stops). In the critical dynamics where the scaling exists, the
corresponding scaling exponents obey the relation 
$\gamma_{sT}=\frac{\tau_T-1}{\tau_s-1}$, where $\gamma_{sT}$ scales
the average size of the avalanches of a given duration $T$, i.e.,
$<s>_T\sim T^{\gamma_{sT}}$.  
In Fig.\ \ref{fig:distributions3x}, we show the results for the
distributions of size and duration of the spin-activity avalanches for the case of purely
pairwise antiferromagnetic coupling, $\kappa=0$, purely triangular
ferromagnetic coupling,$\kappa=1$, and the intermediate case with
balanced pairwise and triangular interactions, $\kappa=0.5$. The
hysteresis loops showing the magnetisation vs external field are given in the corresponding lower panel.  As Fig.\ \ref{fig:distributions3x} shows, the distributions of spin-activity avalanches are exhibiting a power-law decay with an exponential cut-off according to Eq.\
(\ref{eq:Pcsizedur}), however, two sets of scaling exponents apply. 
In particular, when $\kappa=0$, corresponding to strictly pairwise antiferromagnetic
interactions, large avalanches decay with the exponents
$\tau_s-1=0.526\pm 0.018$ and
$\tau_T-1=0.993\pm 0.023$, close to the mean-field SOC \cite{SOC_MF1988cBP}. Whereas, the exponents found by fitting the corresponding
curves in the case of $\kappa=1$ are lower, $\tau_s-1=0.326\pm 0.013$ and
$\tau_T-1=0.49\pm 0.021$. We note that these exponents are numerically close to the ones in the universality class of directed percolation cellular automata \cite{SOC_dirSPA1989DD}. See more discussion below.\\
\begin{figure}[htb]
\begin{tabular}{cc}
\resizebox{24pc}{!}{\includegraphics{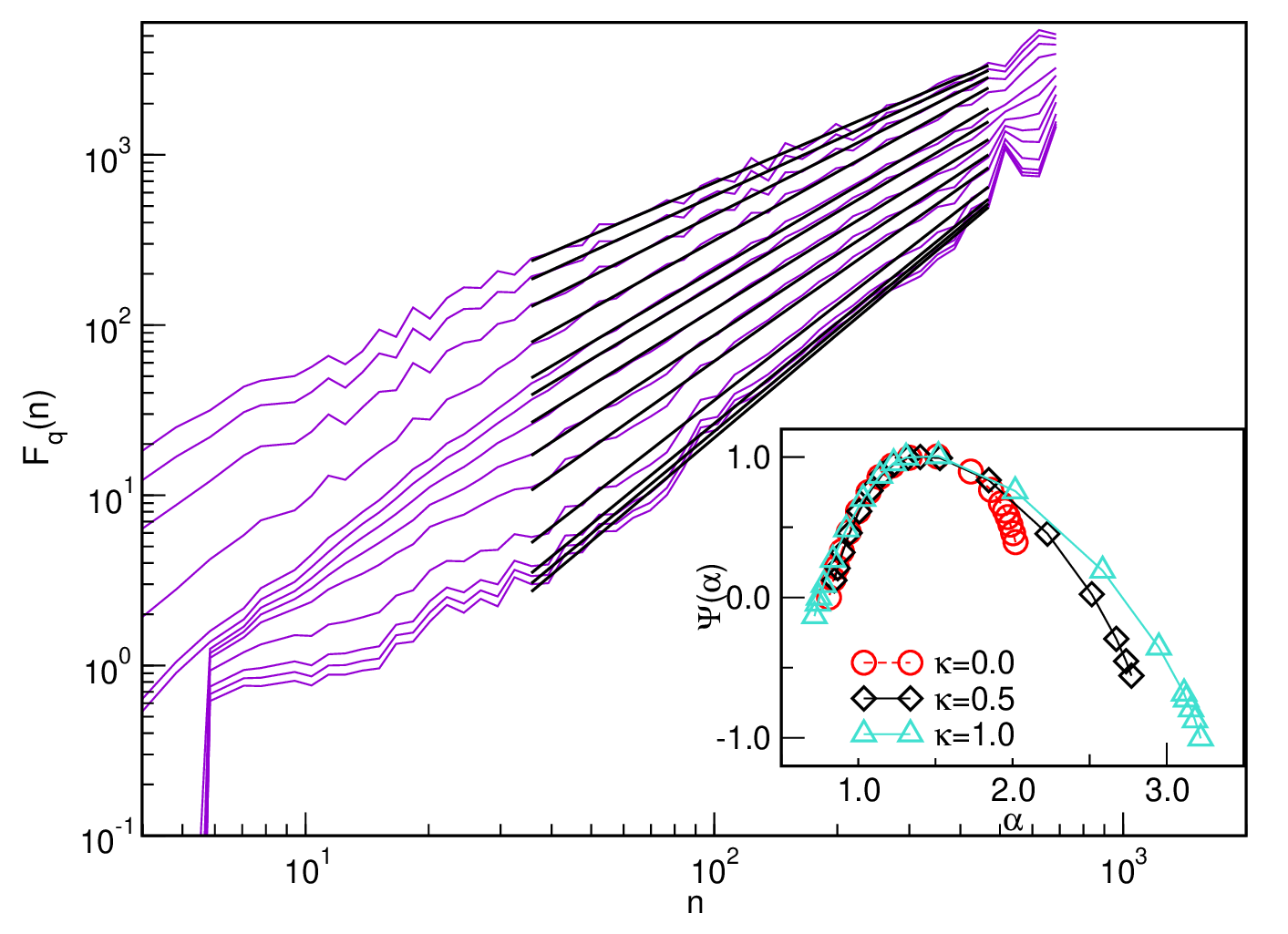}}\\
\end{tabular}
\caption{The fluctuation function $F_q(n)$ vs time interval length $n$
  for  $q\in[-4.5,+4.5]$ (every second line is shown for better vision), for the case $\kappa=0.5$
  of balanced antiferromagnetic pairwise and positive
  triangle-based interactions, using the whole signal up-down loop branch.
 Inset:  the  singularity spectrum $\Psi(\alpha)$ vs $\alpha$ for three cases of the balance pairwise antiferromagnetic and triangle-based
  ferromagnetic interactions, the parameter $\kappa=$0.0,
   $\kappa=$0.5, corresponding to the fluctuation function on the left, and  $\kappa=$1.0.}
\label{fig:Fqupdown_alpha05x3psi}
\end{figure}

To demonstrate the multifractal properties of these magnetisation reversal processes,  we exploit
the underlying scale-invariance of the fluctuation function $F_q(n)$ as a function of the time segment $n$ and determine the
spectrum of generalised Hurst exponents $H_q$. Here, we analyse the time series of the number of spin flips $\{n_t\}$ along the entire hysteresis loop. $T_{max}$ is the time series length.
 In particular, following the procedure for the de-trended multifractal
 analysis described in
 \cite{kantelhardt2002multifractal,we_BHN_MFR2016}, 
the profile $Y(i) =\sum_{t=1}^i(n_t-\langle n_t\rangle) $
of the considered time series is determined. The profile is then divided into $2N_s=2{\rm int}(T_{max}/n)$ non-overlapping segments of the length
$n$, starting from the beginning of the series and then the process is
repeated starting from the end.  
At each segment $\mu=1,2\cdots N_s$, the local trend $y_\mu(i)$ is
determined, and the standard deviation around it is computed as 
 $F(\mu,n) =\left\{ \frac{1}{n}\sum_{i=1}^n[Y((\mu-1)n+i)-y_\mu(i)]^2\right \}^{1/2} $,
and similarly, $F(\mu,n) =\{\frac{1}{n}\sum_{i=1}^n[Y(N-(\mu-N_s)n+i)-y_\mu(i)]^2\}^{1/2}$ for $\mu
=N_s+1,\cdots 2N_s$.
Then, the $r$-th order fluctuation function is for the segment of the
length $n$ is obtained and averaged over all such segments as
\begin{equation}
F_{q}(n)=\left\{\frac{1}{2N_s}\sum_{\mu=1}^{2N_s} \left[F^2(\mu,n)\right]^{ {q}/2}\right\}^{1/ {q}} \sim n^{H_{q}} 
\label{eq:Fq}
\end{equation}
 For varying the parameter $q\in[-4.5,4.5]$, we explore the scale-invariant parts of the fluctuation function $F_q(n)$  against $n$, where the segment length is varied in the range 
$n\in [2,{\rm int}(T_{max}/4)]$, cf.\ Fig.\ \ref{fig:Fqupdown_alpha05x3psi}.
In this way, the segments of the signal representing small and large  fluctuations, respectively, are enhanced in
different ways to maintain the self-similarity of the whole time series.  
Here, a generalised Hurst exponent $H_ {q}$ depends on the
amplification factor $q$, as shown in the inset of Fig.\ \ref{fig:Fqupdown_alpha05x3psi}. Intuitively, this means that the small fluctuations (amplified when $ {q}<0$) along the signal have different scaling properties than the large fluctuations ( {q}$>0$). In contrast, $H_ {q}=H_2$ for all $ {q}$, where $H_2$ is the
standard Hurst exponent for the monofractal signals. 
Having the spectrum of the generalised Hurst exponents $H_q$, we determine other multifractal measures; see, for example, \cite{kantelhardt2002multifractal}. Specifically, we obtain $\tau_q=qH_q-1$, which is related to the standard (box probability) measure, and the singularity spectrum $\Psi(\alpha)=q\alpha -\tau_q$, where $\alpha = d\tau/dq$; in this context, different $\alpha$ values indicate variations in the power-law singularities along the considered time series.  In the inset to Fig.\ \ref{fig:Fqupdown_alpha05x3psi}, we show how the shape of the singularity spectrum changes when the balance between antiferromagnetic pairwise and triangle-embedded interactions is varied via the parameter $\kappa$.

\section{Discussion and future directions\label{sec:discussion}}
In this paper, we have given a survey of recent research activities aiming at understanding the role of higher-order connectivity in functional complex systems. Our leading idea concerns the intricate interdependences of the higher-order structures and the out-of-equilibrium self-organised critical dynamics as a plausible mechanism for emerging new properties at a larger scale, the central paradigm of physical complexity. Having briefly described both issues, we have mentioned an increasing number of studies that evidence the occurrence of higher-order (hidden) structures and signatures of SOC in many complex systems across the scales. We have highlighted the most striking example, mapping the brain structure and its dynamics, which has considerable influence on the science of complexity theory and practice, as well as the applications in artificial intelligence. Furthermore, we highlighted the relevance of different time scales in the light of the structure--dynamics interdependences. In this context, the time-scale separation between the intrinsic dynamics and driving is required for the nonlinear systems to evolve towards SOC attractors. In addition, we stress the relevance of the time scale of changes in the underlying geometry evolution as compared to the SOC dynamics; 
they can be characterised by the appropriate graph or simplicial complex measures. In this regard, we differentiate two limiting cases, particularly structures co-evolving with the intrinsic dynamics and the quenched structure, whose evolution time exceeds both time scales relevant to the SOC. We referred to several pertinent examples of both groups of systems studied in the literature. Another interesting, much less understood case concerns the underlying geometry reconstruction at an intermediate time scale between the intrinsic dynamics and driving. It can occur virtually independent from the dynamics of the actual units comprising the system, for example, reconstructions via built-in temporal defects, or can have a hidden connection with the driving forces. We have briefly described two currently most studied dynamic models---synchronisation of phase oscillators and zero-temperature spin dynamics on simplicial complexes of a fixed architecture, which provides geometry-embedded higher-order interactions. A more detailed study of the spin-reversal dynamics with the competing geometry-embedded interactions leading to SOC behaviour is given as an illustrative example.

In the studied example with Ising spins on an assembly of triangles,
we have shown that the spin reversal zero-temperature dynamics with competing interactions of different orders when driven by avalanches-adapted field changes lead to self-organised critical dynamics on the hysteresis loop. We observe two different universality classes of SOC that are associated with the order of the primary spin interactions. Here, we emphasise that the system with triangle-embedded interactions \textit{without} background pairwise couplings  evolves towards the SOC state in a new universality class, compared to the one induced by the pairwise interactions. 
Specifically, for $\kappa=1$ where only triangle-based interactions are present,  the exponents found by fitting the corresponding
curves are $\tau_s-1=0.326\pm 0.013$, and $\tau_T-1=0.49\pm 0.021$, which are close to the exact exponents of SOC in the directed sandpile automata $\tau_s =4/3$ and $\tau_T=3/2$, derived in \cite{SOC_dirSPA1989DD}. In contrast, numerically determined exponents $\tau_s-1=0.526\pm 0.018$ and
$\tau_T-1=0.993\pm 0.023$ close to the mean-field SOC \cite{SOC_MF1988cBP}, i.e., $\tau_s =3/2$ and $\tau_T=2$,  are found in the limit $\kappa=0$, where only the antiferromagnetic pairwise interactions are present.

To understand the nature of these dynamic critical states, we refer to the theoretical concepts regarding the equivalence between the zero-temperature Ising spin dynamics and the directed percolation cellular automata \cite{CA_DKequivIsing1984prl}. Furthermore, we recall 
that spin-reversal avalanches triggered by the external field change propagate as a directed branching process, where the actual architecture of triangles determines the underlying branching tree. 
In particular, for fractal networks, where the average branching number saturates for significant distances, the study \cite{CN_fractality2007MFST} utilising spanning trees based on the betweenness centrality of edges reveals that the probability of cluster size $s$ (mass when the box length  $\ell_B\to \infty$) scales with the exponent $\tau_s=\gamma/(\gamma-1)$; the exponent $\gamma$ is related to the degree distribution, i.e., the number of edges per node). 
Then we can compute the corresponding duration exponent $\tau_T$ from the scaling relation \cite{SOC_wedirSPAprob} $\tau_s=2-1/\tau_T$, which  gives $\tau_T=(\gamma-1)/(\gamma-2)$.
We note that the generalised degree distribution for the network of finite size, considered in the above numerical simulations,  has an intermediate scale-free segment compatible with the exponent $\gamma\sim 3.8\pm 0.4$, excluding the hub nodes;  meanwhile,  s smaller exponent applies for the low-degree nodes. It is relevant to mention that, in this particular case,  the number of triangles per node scales similarly to the number of edges.
Apart from the structure of potential branches, a crucial difference regarding the nature of avalanche branching occurs due to the spin frustration effects in the limit $\kappa=0$. In this case, each triangle contains at least one frustrated spin, which prevents propagation of the local order set by the current value of the external field. The situation corresponds to a \textit{critical branching process}, where the average branching number remains constant, which leads to the mean-field SOC according to the original work in \cite{SOC_MF1988cBP}. 
On the other side, for triangle-based interactions only ($\kappa=1$), no spin frustration occurs. Multiplicative branching processes are enabled by the network's fractality, leading to the exponents as above, but now the structure is related to the architecture of triangles \cite{SC_HOCweEPL2020}.
 Given the well-known equivalence of the (zero-temperature) Ising model dynamics with the Domany-Kinzel cellular automata for directed percolation \cite{CA_DKequivIsing1984prl}, as well as the related directed sandpile automata with probabilistic toppling \cite{SOC_wedirSPAprob}, we note that the numerical values of the exponents, in this case, are close to the ones of the directed percolation, which are known with more significant numerical precision, i.e., $\tau_s+1.32059$, $\tau_t=1.47244$, and the anisotropy exponent $\zeta=0.63261$); see \cite{SOC_wedirSPAprob} and references there.

In conclusion, the presented survey and the numerically studied example of the hysteresis-loop criticality in a simplicial complex with geometry-embedded interactions highlight the relevance of self-organised dynamics in functional complex systems and the critical role of geometry in the emergence of this type of collective behaviour. 
In this regard, our  results suggest that triangle-embedded interactions play the primary role in critical dynamics beyond standard pairwise interactions in systems with higher-order geometry.
These studies call for more theoretical investigations, i.e.,  by the renormalisation-group methodology, to confirm the existence of different fixed points representing the classes of emergent nonequilibrium critical dynamics and their stability. To this end, equivalent continuous models that consider features of complex geometry (at the local-to-mesoscopic scale) that are relevant to emergent critical dynamics need to be developed, which creates a challenging theoretical problem. We mention some ideas from the physics of hierarchical spin glasses \cite{RG_spinglassParisi2010,RG_CNets2011frontiers}, mapping directed cellular automata onto zero-temperature Ising model \cite{CA_DKequivIsing1984prl}, SOC models with quenched disorder noise with specific correlations \cite{tadic1998disorder,antonovSOC2023}, topological defects and the use of tropical geometry \cite{kalinin2018self} that may serve as an inspiring point in this direction. Another interesting open problem regards the potentials of quantum formalism, such as in reference \cite{CS_emergentQuantumTh2023},  to elucidate the interplay of geometry and time-varying driving forces in the occurrence of self-organised critical states. 
On the other side, several open questions remain accessible to numerical modelling and empirical data analysis using the theoretical concept of self-organised critical behaviour \cite{SOC_we_dynamics2021}. Some examples include the geometry reconstruction associated with the avalanche driving rate, the role of extended defects, and the occurrence of different types of SOC in systems with ``meaningful'' interactions among social subjects in contrast to the physics laboratory systems and others.

One of the critical ingredients of the developed framework that we highlighted and detailed here is the interplay between self-organized critical dynamics and higher-order connectivity networks. Not only does it reveal new lines of theoretical research, but it also offers a fresh perspective for data analysis based on these scientific concepts.
Namely, self-organized criticality as an active mechanism in the evolution of a complex system manifests itself in characteristic patterns of spatiotemporal correlations that a perceptive analysis of data can reveal and thus enable the use of appropriate theory for further predictions (in contrast to considering an ever-increasing amount of data as uncorrelated sets). Such a combined analysis would allow a deeper understanding of the functional characteristics of many complex systems, for example, the brain, which may allow a more critical approach in designing AI algorithms that utilize brain functions and various medical implications.

\textbf{Acknowledgments.}
BT research supported by  the Slovenian
Research Agency the program P1-0044. RM is grateful to the NSERC and the CRC Program for their support, acknowledging also the support of the BERC 2022–2025 program and the Spanish Ministry of Science, Innovation and Universities through the Agencia Estatal de Investigation (AEI) BCAM Severo Ochoa excellence accreditation SEV-2017–0718.

\textbf{Author contribution statement.} Conceptualisation (BT, RM), computation and graphics (BT), data analysis (BT, RM), writing and editing (BT, RM)

\textbf{Data Availability Statement.} The authors declare that the data supporting the findings of this study are available within the paper, and the related references.\\


\begin{thebibliography}{100}

\bibitem{SOC_we_dynamics2021}
Tadi\'c, B., Melnik, R.
\newblock Self-organised critical dynamics as a key to fundamental features of
  complexity in physical, biological, and social networks.
\newblock {\em Dynamics} \textbf{1(2)}, 181--197 (2021)

\bibitem{CS_bookST2018}
Thurner, S., Hanel, R.,  Klimek, P.
\newblock {\em Introduction to the Theory of Complex Systems}.
\newblock Oxford University Press, 2018.

\bibitem{ComplexSystems_review2020}
Gershenson,  C., Siegenfeld, A.~F., Bar-Yam, Y.
\newblock An introduction to complex systems science and its applications.
\newblock {\em Complexity} \textbf{2020}, 6105872 (2020)

\bibitem{ComplexSystems_Estrada2023}
Estrada, E.
\newblock What is a complex system, after all?
\newblock {\em Foundations of Science}, 1572--8471 (2023)

\bibitem{Noextensive_review2023Sym}
Rodríguez, A.,  Pluchino, A.,  Tirnakli, U.,  Rapisarda, A., Tsallis, C.
\newblock Nonextensive footprints in dissipative and conservative dynamical
  systems.
\newblock {\em Symmetry}, \textbf{15(2)}, 444  (2023)

\bibitem{RG_nonequilibrium_UTauber2017}
T\"{a}uber, U.C.
\newblock Phase transitions and scaling in systems far from equilibrium.
\newblock {\em Annual Review of Condensed Matter Physics} \textbf{8(1)}, 185--210  (2017)

\bibitem{RG_envirDyn2023Antonov}
Antonov, A.V.,  Gulitskiy, N.M.,  Kakin, P.I.,   Lebedev, N.M., 
  M. Tumakova, M.M..
\newblock Field-theoretic renormalization group in models of growth processes,
  surface roughening and non-linear diffusion in random environment: Mobilis in
  mobili.
\newblock {\em Symmetry} \textbf{15(8)}, 1556   (2023)

\bibitem{antonovSOC2023}
Antonov, A.V., Kakin, P.I.I, Lebedev, N.M., Luchin, A.Yu.
\newblock Renormalization group analysis of a self-organized critical system:
  intrinsic anisotropy vs random environment.
\newblock {\em J. of Physics A-Mathematical and theoretical} \textbf{56(37)}, 375002 (2023)

\bibitem{CS_emergentQuantumTh2023}
Hubsch, T.,  Minic, D.,  Nikolic, K.,  Pajevic, S.
\newblock On the emergent "quantum" theory in complex adaptive systems.
\newblock {\em arxiv:2310.14100v1}, (2023)

\bibitem{hwa1992avalanches}
Hwa, T.,  Kardar, M.
\newblock Avalanches, hydrodynamics, and discharge events in models of
  sandpiles.
\newblock {\em Physical Review A} \textbf{45(10)}, 7002 (1992)

\bibitem{tadic1998disorder}
Tadi{\'c}, B..
\newblock Disorder-induced critical behavior in driven diffusive systems.
\newblock {\em Physical Review E} \textbf{58(1)}, 168--173  (1998)

\bibitem{Dyn_criticalonNets2008SD}
Dorogovtsev, S.N., Goltsev, A.V., Mendes, J.F.F..
\newblock Critical phenomena in complex networks.
\newblock {\em Rev. Mod. Phys.} \textbf{80}, 1275--1335 (2008)

\bibitem{RG_Networks2018Boguna}
Garc\'ia-P\'erez, G., Bogu\~{n}a, M., Serrano, M.A..
\newblock Multiscale unfolding of real networks by geometric renormalization.
\newblock {\em Nature Phys}, 14:583–589, 2018.

\bibitem{Nets_LowDim2022}
Almagro, P., Bogu\~{n}a, M., Serrano, M.A..
\newblock Detecting the ultra low dimensionality of real networks.
\newblock {\em Nat Commun} \textbf{13}, 6096 (2022)

\bibitem{HOC_review2023boccaletti}
Boccaletti, S., {De Lellis}, P., {del Genio}, C.I., Alfaro-Bittner, K., Criado, R., Jalan, S., Romance, M.
\newblock The structure and dynamics of networks with higher order
  interactions.
\newblock {\em Physics Reports} \textbf{1018}, 1--64 (2023)

\bibitem{HOC_nonlocal2023Kuehn}
 Bick, C.,Gross, E., Harrington, H.A., Schaub, M.T.
\newblock What are higher-order networks?
\newblock {\em SIAM Review} \textbf{65(3)}, 686-731 (2023)


\bibitem{we_PhysA2015}
Andjelkovi\'c, M., Tadi\'c, B., Maleti\'c, S., Rajkovi\'c, M.
\newblock Hierarchical sequencing of online social graphs.
\newblock {\em Physica A: Statistical Mechanics and its Applications} \textbf{436}, 582
  -- 595 (2015)

\bibitem{HOC_evolution2021}
Alvarez-Rodriguez, U., Battiston, G.F., Arruda, G.F., Moreno, Y., Perc, M., Latora, V.
\newblock Evolutionary dynamics of higher-order interactions in social
  networks.
\newblock {\em Nature Human Behaviour} \textbf{5}, 586 - 595  (2021)

\bibitem{SYNCH_Arenas_NatureComm2020}
Skardal, P.S., Arenas, A.
\newblock {Higher order interactions in complex networks of phase oscillators
  promote abrupt synchronization switching}.
\newblock {\em Communications Physics} \textbf{3(1)}, 4200502000485 (2020)

\bibitem{Synch_clustersEVlocaliz2023}
Khanra, P., Ghosh, S., Alfaro-Bittner, K., Kundu, P., 
  Boccaletti, S., Hens, C., Pal, P.
\newblock Identifying symmetries and predicting cluster synchronization in
  complex networks.
\newblock {\em Chaos, Solitons \& Fractals} \textbf{155}, 111703 (2022)



\bibitem{SC_we_Entropy2020}
Tadi\'c. B., Andjelkovi\'c,  M.,\v{S}uvakov, M., Rodgers, G.J..
\newblock Magnetisation processes in geometrically frustrated spin networks
  with self-assembled cliques.
\newblock {\em Entropy}  \textbf{22(3)}, 336  (2020)

\bibitem{OSN_weEntropy2013ABM}
Tadi\'c. B., Gligorijevi\'c, V.,  Mitrovi\'c, M.,  \v{S}uvakov, M.
\newblock Co-evolutionary mechanisms of emotional bursts in online social
  dynamics and networks.
\newblock {\em Entropy} \textbf{15(12)}, 5084--5120 (2013)

\bibitem{dankulov2015dynamics}
Mitrovi{\'c} Dankulov, M.,  Melnik, R., Tadi{\'c}, B.
\newblock The dynamics of meaningful social interactions and the emergence of
  collective knowledge.
\newblock {\em Scientific Reports} textbf{5(1)}, 12197 (2015)

\bibitem{SOC_wePRE2017knowe}
Tadi\'c, B., Mitrovi\'c Dankulov, M.,  Melnik,R.
\newblock Mechanisms of self-organized criticality in social processes of
  knowledge creation.
\newblock {\em Phys. Rev. E} \textbf{96}, 032307 (2017)

\bibitem{SOC_weHBNets2019}
Tadi\'c, B.
\newblock Self-organised criticality and emergent hyperbolic networks:
  blueprint for complexity in social dynamics.
\newblock {\em European Journal of Physics} \textbf{40(2)}, 024002 (2019)

\bibitem{Brain_CNdyn_review2024}
Papo, D., Buld\'u, J.M.
\newblock Does the brain behave like a (complex) network? i. dynamics.
\newblock {\em Physics of Life Reviews} \textbf{48}, 47--98 (2024)

\bibitem{ProgMatt_Exp2016}
Restrepo, D., Mankame, N.D., Zavattieri, P.D.
\newblock Programmable materials based on periodic cellular solids. part i:
  Experiments.
\newblock {\em International Journal of Solids and Structures} \textbf{100-101}, 485 --
  504 (2016)

\bibitem{SC_we_PRE2020}
Tadi\'c, B., \v{S}uvakov, M., 
  Andjelkovi\'c, M., Rodgers, G.J.
\newblock Large-scale influence of defect bonds in geometrically constrained
  self-assembly.
\newblock {\em Phys. Rev. E} \textbf{102}, 032307 (2020)

\bibitem{SOC_Jensen}
Jensen, H.J.
\newblock {\em Self-organized criticality: emergent complex behavior in
  physical and biological systems}.
\newblock Cambridge University Press, Cambridge (1998).

\bibitem{aschwanden2013self}
Aschwanden, M.J.  (Editor)
\newblock Self-organized criticality systems.
\newblock {\em Open Academic Press, Berlin}, 2013.

\bibitem{markovic2014power}
Markovi{\'c}, D., Gros, C.
\newblock Power laws and self-organized criticality in theory and nature.
\newblock {\em Physics Reports} \textbf{536(2)}, 41--74 (2014)

\bibitem{mcateer201625}
McAteer, RT.J., Aschwanden, M.J., Dimitropoulou, M., 
  Georgoulis, M.K., Pruessner, G.,  Morales, L., Ireland,J., 
  Abramenko, V.
\newblock 25 years of self-organized criticality: Numerical detection methods.
\newblock {\em Space Science Reviews} \textbf{198}, 217--266 (2016)

\bibitem{SPA_btw}
Bak, P.,  Tang, C., Wiesenfeld, K.
\newblock Self-organized criticality: An explanation of the 1/f noise.
\newblock {\em Phys. Rev. Lett.} \textbf{59}, 381--384 (1987)

\bibitem{dhar1990self}
Dhar, D.
\newblock Self-organized critical state of sandpile automaton models.
\newblock {\em Physical Review Letters} \textbf{ 64(14)}, 1613 (1990)

\bibitem{bak1995complexity}
Bak, P., Paczuski, M.
\newblock Complexity, contingency, and criticality.
\newblock {\em Proceedings of the National Academy of Sciences} \textbf{92(15)}, 6689--6696 (1995)

\bibitem{wolf2018physical}
Wolf, Y.I., Katsnelson,  M.I., Koonin, E.V.
\newblock Physical foundations of biological complexity.
\newblock {\em Proceedings of the National Academy of Sciences} \textbf{115(37)}, E8678--E8687 (2018)

\bibitem{SOC_rainfalls2015}
Deluca, A., Moloney,  N.R., Corral, {\'A}.
\newblock Data-driven prediction of thresholded time series of rainfall and
  self-organized criticality models.
\newblock {\em Physical Review E} \textbf{91(5)}, 052808 (2015)

\bibitem{smyth2019self}
Smyth, W.D., Nash,  J.D., Moum, J.N.
\newblock Self-organized criticality in geophysical turbulence.
\newblock {\em Scientific Reports} \textbf{9(1)}, 3747 (2019)

\bibitem{SOC_bialek2011}
Mora, T., Bialek, W.
\newblock Are biological systems poised at criticality?
\newblock {\em J. Statistical Physics} \textbf{144(2)}, 268--302 (2011)

\bibitem{SOCliving_Munoz2018review}
Muñoz, M.A.
\newblock Colloquium: Criticality and dynamical scaling in living systems.
\newblock {\em Reviews of Modern Physics} \textbf{90(3)}, 1001 (2018)

\bibitem{SOC_fish2022romu}
Mugica, J., Torrents, J., Cristin, J., Puy, A., Miguel, C., Pastor Satorras, R.
\newblock Scale-free behavioral cascades and effective leadership in schooling
  fish.
\newblock {\em Sci.Rep.} \textbf{12(1)}, 10783 (2022)

\bibitem{SOC_Epi2000}
 Philippe, P.
\newblock Epidemiology and self-organized critical systems: An analysis in
  waiting times and disease heterogeneity.
\newblock {\em Nonlinear Dynamics, Psychology, and Life Sciences} \textbf{4}, 275 --
  295 (2000)

\bibitem{SOC_Epi2014qexpdengue}
Saba,H., Miranda, JVG, Moret, M.A.
\newblock Self-organized critical phenomenon as a q-exponential decay —
  avalanche epidemiology of dengue.
\newblock {\em Physica A: Statistical Mechanics and its Applications} \textbf{413}, 205--211 (2014)

\bibitem{SOC_trafficmanagement2023}
Laval, J.A.
\newblock Self-organized criticality of traffic flow: Implications for
  congestion management technologies.
\newblock {\em Transportation Research Part C: Emerging Technologies} \textbf{149}, 104056 (2023)

\bibitem{SOC_Optimization2018}
Hoffmann, H., Payton, D.W.
\newblock Optimization by self-organized criticality.
\newblock {\em Sci. Rep.} \textbf{8}, 2358 (2018)

\bibitem{SOC_econo2021review}
Tebaldi, C.
\newblock Self-organized criticality in economic fluctuations: The age of
  maturity.
\newblock {\em Front.Phys.} \textbf{8}, 616408  (2021)

\bibitem{we_cycles_emo23}
Tadi\'c, B., {Mitrovi\'c Dankulov}, M.,  Melnik, R.
\newblock Evolving cycles and self-organised criticality in social dynamics.
\newblock {\em Chaos, Solitons\& Fractals} \textbf{171}, 113459 (2023)

\bibitem{Brain_strfunct2023review}
Czime Litwi{\'n}czuk, M., Trujillo-Barreto, N., Muhlert, N., 
  Cloutman, L., Woollams, A.
\newblock Relating cognition to both brain structure and function: A systematic
  review of methods.
\newblock {\em Brain Connectivity} \textbf{13(3)}, 120--132 (2023)

\bibitem{Brain_disorders2023imaging}
Yen, C.,  Lin, C.L., Chiang, M.C..
\newblock Exploring the frontiers of neuroimaging: A review of recent advances
  in understanding brain functioning and disorders.
\newblock {\em Life} \textbf{13(7)},  (2023)

\bibitem{Brain_SOCpsychedelic2023}
Hip\'olito, I., Mago, J., Rosas, F.E.,  Carhart-Harris, R.
\newblock {Pattern breaking: a complex systems approach to psychedelic
  medicine}.
\newblock {\em Neuroscience of Consciousness} \textbf{2023(1)}, niad017 (2023)



\bibitem{Brain_AI2022dynmodels}
Ramezanian-Panahi, M., Abrevaya, G., Gagnon-Audet, J-C.,
  Voleti, IV., Rish, I.,  Dumas, G.
\newblock Generative models of brain dynamics.
\newblock {\em Frontiers in Artificial Intelligence} \textbf{5}, 807406 (2022)

\bibitem{Brain_horizons2022review}
Srivastava, P., Fotiadis, P., Parkes, L., Bassett, D.S.
\newblock The expanding horizons of network neuroscience: From description to
  prediction and control.
\newblock {\em NeuroImage} \textbf{258}, 119250 (2022)

\bibitem{Brain_SOC2014Thilo}
Hesse, J., Thilo Gross, T.
\newblock Self-organized criticality as a fundamental property of neural
  systems.
\newblock {\em Frontiers in Systems Neuroscience} \textbf{8},  00166 (2014)

\bibitem{gros2021devil}
Gros, C.
\newblock A devil’s advocate view on ‘self-organized’brain criticality.
\newblock {\em Journal of Physics: Complexity} \textbf{2(3)}, 031001  (2021)

\bibitem{Brain_SOCexp2021review}
Plenz, D., Ribeiro, T.L., Miller, S.R., Kells, P.A.,
  Vakili, A., Capek, E.L..
\newblock Self-organized criticality in the brain.
\newblock {\em Frontiers in Physics} \textbf{9},  639389 (2021)

\bibitem{Brain_Connectomics2017review}
Lord, L-D., Stevner, A.B., Deco, G., Kringelbach, M.L.
\newblock Understanding principles of integration and segregation using
  whole-brain computational connectomics: implications for neuropsychiatric
  disorders.
\newblock {\em Physlosophical Transactions A} \textbf{375}, 20160283 (2017)


\bibitem{Brain_intseg2021attention}
Zuberer, A., Kucyi, A., Yamashita, A., Wu, C.M., Walter, M.,
  Valera, E.M., Esterman, M.
\newblock Integration and segregation across large-scale intrinsic brain
  networks as a marker of sustained attention and task-unrelated thought.
\newblock {\em NeuroImage} \textbf{229}, 117610  (2021)

\bibitem{Brain_intsegEEG2018cognition}
Cruzat, J., Deco, G., Tauste-Campo, A., Principe, A., 
   Costa, A., Kringelbach, M.L., Rocamora, R.
\newblock The dynamics of human cognition: Increasing global integration
  coupled with decreasing segregation found using ieeg.
\newblock {\em NeuroImage} \textbf{172}, 492--505 (2018)

\bibitem{Brain_intseg2022pain}
Kastrati, G., Thompson,  W.H., Schiffler, B., Fransson, P., 
  Jensen, C.B..
\newblock {Brain network segregation and integration during painful thermal
  stimulation}.
\newblock {\em Cerebral Cortex} \textbf{32(18)}, 4039--4049 (2022)

\bibitem{Brain_Nets2023review}
Seguin, C., Sporns, O., Zalesky, A.
\newblock Brain network communication: concepts, models and applications.
\newblock {\em Nature Review Neurocsience} \textbf{24}, 557--574 (2023)

\bibitem{Brain_synch2021psychiatry}
Anyaeji, CI., Cabral, J., Silbersweig, D.
\newblock On a quantitative approach to clinical neuroscience in psychiatry:
  Lessons from the Kuramoto model.
\newblock {\em Harv Rev Psychiatry} \textbf{29(4)},  318-326 (2021)

\bibitem{Brain_Synch2023PLOScbiol}
 Mackay, M., Huo, S., Kaiser, M.
\newblock Spatial organisation of the mesoscale connectome: A feature
  influencing synchrony and metastability of network dynamics.
\newblock {\em PLOS Computational Biology} \textbf{19(8)}, 1--18 (2023)

\bibitem{Synch_noBrainsyncro2019}
Papo, D.,  Buld\'u, J.M.
\newblock Brain synchronizability, a false friend.
\newblock {\em NeuroImage} \textbf{196}, 195--199 (2019)

\bibitem{we_Braindyn2022ChSFR}
Tadi\'c, B.,  Chutani, M., Gupte, N.
\newblock Multiscale fractality in partial phase synchronisation on simplicial
  complexes around brain hubs.
\newblock {\em Chaos, Solitons\& Fractals} \textbf{160}, 112201 (2022)

\bibitem{Nets_3xBoguna2021}
Bogu\~{n}a, M., Bonamassa, I., De~Domenico, M.,  Havlin,  S., Krioukov, D., 
  Serrano, M.A.
\newblock Network geometry.
\newblock {\em Nature Reviews Physics} \textbf{3(2)},  114 -- 135  (2021)

\bibitem{sergey-lectures}
Dorogovtsev, S.
\newblock {\em Lectures on Complex Networks}.
\newblock Oxford University Press, Inc., New York, NY, USA, 2010.

\bibitem{kozlov-book}
Kozlov D.
\newblock {\em Combinatorial Algebraic Topology}.
\newblock Springer Series Algorithms and Computation in Mathematics , Vol. 21,
  Springer-Verlag Berlin Heidelberg, 2008.


\bibitem{jj-book}
Jonsson J.
\newblock {\em Simplicial Complexes of Graphs}.
\newblock Lecture Notes in Mathematics, Springer-Verlag, Berlin, 2008.

\bibitem{we-Brain_SciRep2019}
Tadi\'c, B., Andjelkovi\'c, M., Melnik, R.
\newblock Functional geometry of human connectomes.
\newblock {\em Scientific Reports} \textbf{ 9}, 12060 (2019)

\bibitem{HOC_neuronsnets2019}
Tlaie, A., Leyva, I., Sendinna-Nadal, I.
\newblock Higher-order couplings in geometric complex networks of neurons.
\newblock {\em Phys. Rev. Lett.} \textbf{100}, 052305 (2019)

\bibitem{HC_cliquecavities_JCompNeurosci2018}
Sizemore, A.E., Giusti, C., Kahn, A., Vettel, J.M.
\newblock Cliques and cavities in human connectome.
\newblock {\em J Comput. Neurosci.} \textbf{44}, 115--145 (2018)

\bibitem{we-Brain_SciRep2020}
Andjelkovi\'c, M., Tadi\'c, B., Melnik, R.
\newblock The topology of higher-order complexes associated with brain hubs in
  human connectomes.
\newblock {\em Scientific Reports} \textbf{ 10}, 17320 (2020)

\bibitem{AT_Materials_Jap2016}
Ikeda,~S. Kotani,~M.
\newblock Materials inspired by mathematics.
\newblock {\em Science and Technology of Advanced Materials} \textbf{17(1)}, 253--259 (2016)

\bibitem{AT_cooperativeSA2021}
Samaresh, S., Raval, P., {G.N. Manjurata}, R., Debangshu, C.
\newblock Cooperative self-assembly driven by multiple noncovalent
  interactions: Investigating molecular origin and reassessing
  characterization.
\newblock {\em ACS Cent. Sci.} \textbf{7(8)},  1391 --1399   (2021)

\bibitem{shapoval2021predictability}
Shapoval, A.,  Savostianova, D., Shnirman, M.
\newblock Predictability and scaling in a btw sandpile on a self-similar
  lattice.
\newblock {\em Journal of Statistical Physics} \textbf{ 183(1)}, 14 (2021)

\bibitem{HOC_control2023ChSFR}
Qiu, X., Yang, L., Guan, C., Leng, S.
\newblock Closed-loop control of higher-order complex networks: Finite-time and
  pinning strategies.
\newblock {\em Chaos, Solitons\& Fractrals} \textbf{173}, 113677 (2023)

\bibitem{SC_we_SciRep2018}
\v{S}uvakov, M., Andjelkovi\'c, M., Tadi\'c, B.
\newblock Hidden geometries in networks arising from cooperative self-assembly.
\newblock {\em Scientific Reports} \textbf{8}, 1987 (2018)

\bibitem{SC_flavorquantum2016GB}
Bianconi, G., Rahmede, C.
\newblock Network geometry with flavor: From complexity to quantum geometry.
\newblock {\em Phys. Rev. E} \textbf{93}, 032315 (2016)

\bibitem{SC_Sfboccaletti2021}
Kovalenko, K.,  Sendina-Nadal, I.,  Khalil, N., Dainiak, A., Musatov, D.,  Raigorodskii, A.M., Alfaro-Bittner, K.,  Barzel, B.,  Boccaletti, S. 
\newblock Growing scale-free simplices.
\newblock {\em Communications physics} \textbf{4}, 43 (2021)

\bibitem{Q-analysis-book1982}
Beaumont, J.R., Gatrell, A.C.
\newblock {\em An Introduction to Q-Analysis}.
\newblock Geo Abstracts, Norwich-Printed by Edmund Nome Press, Norwich, 1982.

\bibitem{Spectra_Nets2003SD}
Dorogovtsev, S.~N., Goltsev, A.~V.,  Mendes,  J.~F.~F.,  Samukhin, A.~N.
\newblock Spectra of complex networks.
\newblock {\em Phys. Rev. E} \textbf{68}, 046109 (2003)

\bibitem{Spectra_dsSynch2019GB}
Mill\'an, A.P., Torres, J.J., Bianconi, G.
\newblock Synchronization in network geometries with finite spectral dimension.
\newblock {\em Phys. Rev. E} \textbf{99}, 022307 (2019)

\bibitem{Spectra_wePRE2019SC}
Mitrovi\'c Dankulov, M., 
  Tadi\'c, B., Melnik, R.
\newblock Spectral properties of hyperbolic nanonetworks with tunable
  aggregation of simplexes.
\newblock {\em Phys. Rev. E}  \textbf{100}, 012309 (2019)

\bibitem{HB_cliques2017plus1}
Cohen, N., Coudert, D., Ducoffe, G., Lancin, A.
\newblock Applying clique-decomposition for computing gromov hyperbolicity.
\newblock {\em Theoretical Computer Science} \textbf{690}, 114--139 (2017)

\bibitem{SC_HOCweEPL2020}
Tadi\'c, B., Gupte, N.
\newblock Hidden geometry and dynamics of complex networks: Spin reversal in
  nanoassemblies with pairwise and triangle-based interactions.
\newblock {\em Europhysics Letters} \textbf{132(6)}, 60008 (2021)

\bibitem{SYNC_globalGinestraPRL23}
Carletti, T., Giambagli, L., Bianconi, G.
\newblock Global topological synchronization on simplicial and cell complexes.
\newblock {\em Phys. Rev. Lett.} \textbf{130}, 187401 (2023)

\bibitem{SC_synchro_wePRE2021}
Chutani, M., Tadi\'c, B.,  Gupte, N.
\newblock Hysteresis and synchronization processes of kuramoto oscillators on
  high-dimensional simplicial complexes with competing simplex-encoded
  couplings.
\newblock {\em Phys. Rev. E} \textbf{104}, 034206 (2021)

\bibitem{SC_synchro_wePRE2023}
Sahoo, S., Tadi\'c, B., Chutani, M.,  Gupte, N.
\newblock Effect of hidden geometry and higher-order interactions on the
  synchronization and hysteresis behavior of phase oscillators on 5-clique
  simplicial assemblies.
\newblock {\em Phys. Rev. E} \textbf{108}, 034309 (2023)

\bibitem{Synch_groupsPRE2013experiment}
Williams, C. R.S., Murphy, T.E., Roy, R., Sorrentino, F., 
  Dahms, T., Sch\"oll, E.
\newblock Experimental observations of group synchrony in a system of chaotic
  optoelectronic oscillators.
\newblock {\em Phys. Rev. Lett.} \textbf{110}, 064104 (2013)

\bibitem{Synch_clusters2020NatComm}
Della~Rossa, F., Pecora, L., Blaha, K., Shirin, A., Klickstein, I., 
   Sorrentino, F.
\newblock Symmetries and cluster synchronization in multilayer networks.
\newblock {\em Nature communications} \textbf{11(1)}, 3179 (2020)

\bibitem{tadic2019critical}
Tadi{\'c}, B., Mijatovi{\'c}, S., Jani{\'c}evi{\'c}, S., 
  Spasojevi{\'c}, D., Rodgers, G.J.
\newblock The critical barkhausen avalanches in thin random-field ferromagnets
  with an open boundary.
\newblock {\em Scientific reports} \textbf{9(1)}, 6340 (2019)

\bibitem{Frustr_book2015Julich}
Mila, F.
\newblock {\em Frustrated Spin Systems}.
\newblock E. Pavarini, E. Koch, and P. Coleman (eds.) Many-Body Physics: From
  Kondo to Hubbard Modeling and Simulation Vol. 5 Forschungszentrum J\"ulich,
  2015, ISBN 978-3-95806-074-6, 2015.

\bibitem{SOC_MF1988cBP}
Alstr{\o}m, P.
\newblock Mean-field exponents for self-organized critical phenomena.
\newblock {\em Phys. Rev. A} \textbf{38}, 4905--4906  (1988)

\bibitem{SOC_dirSPA1989DD}
Dhar, D., Ramaswamy, R.
\newblock Exactly solved model of self-organized critical phenomena.
\newblock {\em Phys. Rev. Lett.} \textbf{63}, 1659--1662 (1989)

\bibitem{kantelhardt2002multifractal}
Kantelhardt, J.W., Zschiegner, S.A., Koscielny-Bunde, E.,  Havlin, S., 
  Bunde, A., Stanley, H.E.
\newblock Multifractal detrended fluctuation analysis of nonstationary time
  series.
\newblock {\em Physica A: Statistical Mechanics and its Applications}
  \textbf{316(1-4)}, 87--114 (2002)

\bibitem{we_BHN_MFR2016}
Tadi{\'c}, B.
\newblock {Multifractal analysis of Barkhausen noise reveals the dynamic nature
  of criticality at hysteresis loop}.
\newblock {\em Journal of Statistical Mechanics: Theory and Experiment}
  \textbf{6(6)}. 063305 (2016)

\bibitem{CN_fractality2007MFST}
Kim, K-I., Goh, J.S., Salvi, G., Oh, E., Kahng, B., ~Kim, D.
\newblock Fractality in complex networks: Critical and supercritical skeleton.
\newblock {\em Phys. Rev. E} \textbf{75}, 016110 (2007)

\bibitem{SOC_wedirSPAprob}
Tadi\'c, B., Dhar, D.
\newblock Emergent spatial structures in critical sandpiles.
\newblock {\em Phys. Rev. Lett.}  \textbf{79}, 1519--1522 (1997)



\bibitem{CA_DKequivIsing1984prl}
Domany, E., Kinzel, W.
\newblock Equivalence of cellular automata to ising models and directed
  percolation.
\newblock {\em Phys.Rev. Lett.} \textbf{53}, 311--314 (1984)

\bibitem{RG_spinglassParisi2010}
Castellana, M., Parisi, G.
\newblock Renormalization group computation of the critical exponents of
  hierarchical spin glasses.
\newblock {\em Phys. Rev. E} \textbf{82}, 040105 (2010)

\bibitem{RG_CNets2011frontiers}
Boettcher, S.
\newblock Renormalization group for critical phenomena in complex networks.
\newblock {\em Frontiers in Physiology} \textbf{2},  (2011)

\bibitem{kalinin2018self}
Kalinin, N., Guzm{\'a}n-S{\'a}enz, A., Prieto, Y., Shkolnikov, M., 
  Kalinina, V., Lupercio, E.
\newblock Self-organized criticality and pattern emergence through the lens of
  tropical geometry.
\newblock {\em Proceedings of the National Academy of Sciences} \textbf{115(35)}, E8135--E8142 (2018)

\end{thebibliography}

\end{document}